\documentclass{article}

\begin{document}

\author{R.V. TUMANIAN\\,
Yerevan Physics Institute,Armenia\\
E-mail: raphael@star.yerphi.am}
\title{RESONANT LASER COOLING OF CIRCULAR ACCELERATOR BEAMS}
\maketitle
\begin{abstract}

The resonant laser cooling of circular accelerator beams of relativistic charged 
particle is studied. It is shown that in the  approximation of the given 
external electromagnetic wave amplitude (small gain free electron laser) the 
emittance of a beam of charged particles decreases. In the range of particle energy 
about 100 in the mass energy units the beam energy losses are negligible. 
The discovered effect can be used for cooling of charged particle beams in 
various accelerators. The significant cooling rates are possible to achieve by 
placing of the cooling device in the appropriate points of the accelerator orbit. 
Resonant laser cooling for various cases are considered. This method of 
charged particle beams cooling can be applied to circular accelerators of 
electron and various relativistic heavy particle beams and has significantly, 
about 3-4 orders of magnitude, shorter cooling time in comparison to any other 
cooling method.
\end{abstract}
\begin{flushleft}
\begin{twocolumn}

\section{Introduction}

 The increasing of the accelerator beam lifetime and luminosity is very important. 
But the beam lifetime and emmitance is restricted by various effects. For example, 
in the electron cyrcular accelerators the beam lifetime is restricted by synchrotron 
radiation, especially in the machines with combined function magnets as 
ARUS (Yerevan) and old DESY (Germany). In the proton and ion cyrcular accelerators 
the beam lifetime and emmitance is restricted by the different scattering effects. 
The influence of these effects on the beam lifetime and emittance can be decreased by 
applying of any effective cooling method. So, the cooling of charged particle 
beams is a very important issue for attaining of higher lifetime and luminosity beams. 
To date, variuos methods of cooling are used in various accelerators: there are 
radiative \cite{1}, electron \cite{2}, stochastic  \cite{3} and Doppler  \cite{4} cooling. 
The first method is mainly used to cool electron or light particle beams in circular 
accelerators and storage rings. Electron cooling proposed by Budker \cite{2} is
advantageous to cool proton and heavy particle beams.  Exellent reviews on  
cooling of charged particle beams are given in  \cite{2a}-\cite{2b1}. 
Stochastic cooling is applicable to various particle 
beams, but it is very slow. Doppler cooling is used for ion beams.  In \cite{4b}
laser cooling of relativistic ion beam is observed. The lowest 
temperature of a laser cooled stored ion beam is reported in \cite{4c}. 
However, above methods do not meet the modern requirements.
For instance, these methods do not provide the high luminosity beams  
in linear, circular accelerators 
and storage rings. Also they have small cooling rate resulting in long cooling time. 

Moreover, there are no effective cooling methods for high energy, relativistic 
particle beams used in linacs and colliders. In order to
improve  the cooling efficiency new methods have been proposed,
among which we note a resonant laser  \cite{5,5a}, an undulator  \cite{6}, 
ionization  cooling \cite{7}, the stimulated 
radiation or stimulated Doppler cooling \cite{9}, and 
the 3-dimensional laser cooling \cite{8}. 

Transverse laser cooling of ions is demonstrated in the \cite{8a}. Cooling by backward 
Compton scattering of free electrons on photons  is of a special interest \cite{10}-\cite{12}. 
It is known, however, that the stimulated scattering under 
the same conditions has a greater cross-section \cite{9}, \cite{13}, than the 
cross-section  of free particle scattering. Therefore, in contrast to cooling 
methods depicted above, it is advantageous to consider stimulated or resonant 
scattering. In this paper we discuss the applicability of stimulated or 
resonant interaction of particles with a flat electromagnetic wave (photons)
for fast transverse cooling of a beam of relativistic charged particles. 
The resonant bunch - wave interaction is mediated by applying the homogeneous
magnetic field, directed along of a beam movement direction.
The magnetic field creates beam current density correlations proportional to laser 
wavelength. This increases bunch - wave interaction and makes it selective.
The usual and known distribution functions, for example Gaussian or Boltzman, 
are useful for ensembles of particles with irregular, stochastic trajectories, when 
the exact trajectories are not known and the ensemble is described by the density of 
probability of particle states. When particles are moved at certain trajectories,
i.e. we know the particle state, the distribution is not purely probabilistic, rather 
it becomes correllated. This can be seen from the distribution function of 
particles with certain trajectories, which is equal to usual distribution function for 
random variables, for example momentum, after substituting exact trajectories 
instead of random variables. Since the trajectories depend on time, the distribution 
function depends on time too. This show that temporary evolution of the ensemble 
is not probabilistic, and is determined by the certainity of trajectories. All average 
macroscopic values of the system depend on time and we know 
exactly their temporal evolution, which means that correlations of states of 
particles exist in such system.

\section{Physical principles of resonant laser cooling}
\subsection{Introduction}
Interaction of electrons, which move by  periodic trajectories,  with a flat 
monochromatic electromagnetic wave is one way of amplification of an 
electromagnetic wave in the free electron laser. The  autoresonant 
free electron laser \cite{14} based on the interaction of the charged particle with the 
electromagnetic wave extending along a homogeneous magnetic field 
$B_0 \parallel z$, is characterized by the fact that detuning  from an exact 
resonance stays independent on charged particle energy changes. In \cite{15} it 
 was shown , that detuning is constant due to the existence of the  integral 
 of motion $I$
\begin{equation}
I = \gamma  - P_z,
\label{intmot}
\end{equation}
which is applicable only in a constant external wave approximation. Here 
$\gamma$ is the full energy, and $P_z$ is the momentum of a particle along 
magnetic field (in the system of units with m=c=1). This formula can be obtained 
easily from the usual hamiltonian of a charged particlle in the presence of a 
plane laser wave proportional to $exp{i(\omega t-kz)}$. The dependence on time 
means that energy is not conserved and hamiltonian is not a motion integral. 
The new motion integral can be found  from the Hamiltonian by using the 
canonical transformation $\phi=t-z$. After such transformation 
of z, the new hamiltonian equals to I and do not depend on time. 
The time independence of the new Hamiltonian means that new energy I is 
conserved and I is a motion integral. 
In this article we examine the influence of autoresonant interaction of a particle 
with a wave on a transverse beam emittance. Taking into account 
the expression $\gamma  = \sqrt {P_z^2  + P_ \bot ^2  + 1} $,  it is easy to 
receive from (1) for a transverse kinematical momentum of a particle
\[
P_ \bot ^2  = 2I\gamma  - 1 - I^2 
\]
After differentiating of this expression with respect to time and taking an average 
of all particles in a beam, we shall receive the equation for changing of the 
mean square of a transverse momentum of a beam
\begin{equation}
\frac{d}{{dt}}\langle P_ \bot ^2 \rangle  = 2\langle I\dot \gamma \rangle
\label{eqmot} 
\end{equation}
As shown in works \cite{14}-\cite{16}, the trajectories of electrons are 
spirals, which radii  varyes along the way. As the drift of the center 
of a Larmor circle is absent and $I \geq 0$, 
interaction in the mode of the laser ($\dot \gamma  < 0$) means reduction of a 
beam transverse emittance, because of the negativity of the right hand side of (\ref{eqmot}). 
Here and below the dot means the differentiating on time. 
In the laser mode the amplified 
wave carries out a part of beam energy. 

It is very important to find 
which part of the energy is taken out from full beam energy and 
which part is taken from beam transverse emittance, i.e. what constitutes beam cooling 
(emittance reduction). As it is not difficult to receive from \cite{intmot}, relative speed 
$\dot P_z /P_z $ of a longitudinal momentum change is $2P_ \bot ^2 /(1 + P_ \bot ^2 )$ 
times less than relative speed $\dot P_ \bot  /P_ \bot  $ of a transverse 
momentum change. Therefore emittance reduction will be effective for beams with 
$\langle P_ \bot ^2 \rangle  \ll 1$, when full energy losses are small. But in 
practical applications it is also possible to compensate beam energy losses 
by external accelerating elements. 

\subsection{Equations of motion}
In the limit of infinitesimal changes of a transverse momentum during 
the beam interaction with a wave, i.e. if motion in a transverse plane is determined 
by a magnetic field $B_0 $, it is easy to receive for $\dot \gamma $ in the 
field of the circular-polarized wave with frequency $\omega$ and amplitude Å 	
\[
\dot \gamma  = \xi \omega \frac{{P_ \bot  }}{\gamma }\cos \varphi 
\]
where $\xi  = {\raise0.7ex\hbox{${eE}$} \!\mathord{\left/
 {\vphantom {{eE} \omega }}\right.\kern-\nulldelimiterspace}
\!\lower0.7ex\hbox{$\omega $}}$ is the dimensionless amplitude of an electric field of 
a wave, and $\varphi$ is the phase of the Larmor rotation, counted from 
a wave phase at the location of a particle. In the approximation used it 
is easy to derive the equation for the change of an average longitudinal momentum 
or energy of a beam
\begin{equation}
\left\langle {\dot p} \right\rangle  = \left\langle {\dot \gamma } \right\rangle  = \left\langle {\xi \omega \frac{{p_ \bot  }}{\gamma }\cos \varphi } \right\rangle 
\label{longmom}
\end{equation}
Averaging on all particles of a beam in the equation (\ref{eqmot}) allows to investigate 
the average beam size, using the kinetic approach instead of particle-by-particle  
detailed trajectory analysis.  If a weak (in comparison to the magnetic field $\vec B_0 $) 
electromagnetic wave with $\vec E,\vec B \sim exp{i(\omega t - kz)} $ is present,
there are correlations of density which result in modulation of a 
beam on $\varphi$ and accordingly, give a nonzero right hand side in (\ref{eqmot}). 
To quantify this effect we shall present distribution function of a beam as  \cite{17}
$f = f_0  + \delta f$, where $f_0 $ is the basic equilibrium function of distribution, 
and $\left| {\delta f} \right| \ll f_0 $ is the small addition to $f_0 $ caused by 
a wave. Close to a resonance it is convenient to write the addition as 
Fourier series 
$\varphi :\delta f = \sum\limits_{}^{} {g_s (P_ \bot  ,P_z )exp{(is\varphi) } } $. 
Substituting this in the kinetic equation and neglecting terms of the second 
order, gives the following expressions for factors
\[
g_s  = \frac{{Q_s }}{{i(s + \alpha )}};Q_s  = \frac{1}{{2\pi }}\int\limits_0^{2\pi } {d\tau exp{( - is\tau) } Q(P_z ,P_ \bot  ,\tau );Q = \frac{e}{{\omega _B }}\frac{{\partial f_0 }}{{\partial \vec P}}(\vec E + \frac{1}{\omega }[\vec v[\vec k\vec E]])} 
\]
where $\alpha  = {\raise0.7ex\hbox{${(kv_z  - \omega )}$} \!\mathord{\left/
 {\vphantom {{(kv_z  - \omega )} {\omega _B }}}\right.\kern-\nulldelimiterspace}
\!\lower0.7ex\hbox{${\omega _B }$}}$,  
$\omega _B  = {\raise0.7ex\hbox{${eB_0 }$} \!\mathord{\left/
 {\vphantom {{eB_0 } \gamma }}\right.\kern-\nulldelimiterspace}
\!\lower0.7ex\hbox{$\gamma $}}$  is the Larmore frequency, $\tau$ is the phase 
of integration and the relation $\omega \vec B = [\vec k\vec E]$  for the field 
of the wave is used. Close to a simple cyclotron resonance the leading term in the 
Fourier series will be the term with s=1. For Gaussian distribution function 
on  transverse momentum 
\begin{equation}
f_0  = \frac{1}{{\pi \left\langle {P_ \bot ^2 } \right\rangle }}exp{( - {\raise0.7ex\hbox{${P_ \bot ^2 }$} \!\mathord{\left/
 {\vphantom {{P_ \bot ^2 } {\left\langle {P_ \bot ^2 } \right\rangle }}}\right.\kern-\nulldelimiterspace}
\!\lower0.7ex\hbox{${\left\langle {P_ \bot ^2 } \right\rangle }$}})} \delta (P_z  - P_0 ).
\label{f}
\end{equation}
In a relativistic case $P_0  \gg 1 + P_ \bot ^2 $ we can neglect the $I^2$ in (1) 
and taking into account $I \approx {{(1 + P_ \bot ^2 )} \mathord{\left/
 {\vphantom {{(1 + P_ \bot ^2 )} {2P_0 }}} \right.
 \kern-\nulldelimiterspace} {2P_0 }}$ 
we find a small addition to (\ref{f})
\begin{equation}
\delta f =  - \xi \frac{{P_ \bot  (1 + P_ \bot ^2 )}}{{\left\langle {P_ \bot ^2 } \right\rangle i(\Delta _\parallel   - P_ \bot ^2 )}}exp{(i\varphi) } f_0 
\end{equation}
 where
$\Delta _\parallel   = 2\frac{\Omega }{\omega }P_0  - 1;\Omega  = \omega _B \gamma $. 
Approximation of a longitudinal - monochromatic beam $f_0 $ is fair, if the relative 
spread on a longitudinal momentum $(P_z  - P_o )/P_0 $ is much less than 
$\left\langle {P_ \bot ^2 } \right\rangle $. This approximation is quite useful as 
it allows us to estimate $f_0 $ by a resonant term $(1 + \alpha )^{ - 1} $. Then we 
average on f we note rather usual in 
such cases pole located at  $p_ \bot ^2  = \Delta _\parallel  $, which is caused by a 
resonant multiplier. By a rule of detour of Landau poles  (substitution
$\omega  \to \omega  + i0$) we find the vale of the integral of interest \cite{17}: 

\[
\int\limits_0^\infty  {\frac{{f(z)}}{{\Delta _\parallel   - z}}} dz = \int {\frac{{f(z)}}{{\Delta _\parallel   - z}}dz}  + i\pi \delta (\Delta _\parallel   - z)
\]
where $z = p_ \bot ^2 $, and the integral in the right hand side is treated in sense of 
a principal value. As  only a real part of (\ref{eqmot}) is physical  meaningful, if
$\Delta _\parallel   \le 0$, when the right hand side is pure imaginary, the 
speed of change of a 
transverse momentum is equal to zero. And when $\Delta_\parallel > 0$ we find: 

\begin{equation}
\frac{d}{{dt}}\left\langle {p_ \bot ^2 } \right\rangle  =  - \xi \frac{{\pi \omega }}{{2p_0^2 \left\langle {p_ \bot ^2 } \right\rangle }}\Delta _\parallel  (1 + \Delta _\parallel  )exp{( - {{\Delta _\parallel  } \mathord{\left/
 {\vphantom {{\Delta _\parallel  } {\left\langle {p_ \bot ^2 } \right\rangle }}} \right.
 \kern-\nulldelimiterspace} {\left\langle {p_ \bot ^2 } \right\rangle }})} 
 \label{basequat}
\end{equation}
\subsection{Particular solutions}
The maximum of the right hand side in case with $\left\langle {p_ \bot ^2 } \right\rangle  \ll 1$
($p_0  \approx const$) is achieved at
$\Delta _\parallel   = \left\langle {p_ \bot ^2 } \right\rangle $, and its value is
\begin{equation}
\left\langle {p_ \bot ^2 } \right\rangle  = (\left\langle {p_{ \bot 0}^2 } \right\rangle ^2  - \frac{{2\pi ^2 }}{e}\xi ^2 \frac{z}{{2\lambda p_0^2 }})^{1/2} 
\label{A1}
\end{equation}
where $P_{\bot 0}$ is the initial value of an average transverse 
momentum of a beam squared, 
($\lambda$ is a wavelength, and z =ct is a length of a beam way. When 
$\Delta _\parallel   = const$ 
and much less $\left\langle {p_ \bot ^2 } \right\rangle $, we shall receive: 
\begin{equation}
\left\langle {p_ \bot ^2 } \right\rangle  = (\left\langle {p_{ \bot 0}^2 } \right\rangle ^3  - \xi ^2 \frac{{6\pi ^2 z}}{{2\lambda p_0^2 }}\Delta _\parallel  )^{1/3} 
\label{A2}
\end{equation}

The decreasing of a beam transverse emittance by autoresonant bunch - wave interaction 
must be significant as well for a bunch modulated $\varphi$. In this case 
we choose the distribution function of a different kind: 
\[
f(p_ \bot  ,p_z ,\varphi ) = f_0 (1 - \varepsilon \cos \varphi )
\]
where $0 < \varepsilon  \le 1$ is the depth of modulation, and $f_0 $ is defined 
in (\ref{f}). After averaging of (2) with this function we shall find: 
\begin{equation}
\frac{d}{{dt}}\left\langle {p_ \bot  } \right\rangle  =  - \varepsilon \xi \frac{\omega }{{2p_0^2 }}(1 + \frac{6}{\pi }\left\langle {p_ \bot  } \right\rangle ^2 ),
\end{equation}
where
$\left\langle {p_ \bot  } \right\rangle  = \frac{{\sqrt \pi  }}{2}\sqrt {\left\langle {p_ \bot ^2 } \right\rangle } $
relation is taken into account. From here at 
$\left\langle {p_ \bot ^2 } \right\rangle  \ll 1$ ($p_0  \approx const$), we shall 
obtain 
	
\begin{equation}
\left\langle {p_ \bot  } \right\rangle  = p_{ \bot 0}  - \varepsilon \xi \pi \frac{z}{{2\lambda p_0^2 }}
\label{A3}
\end{equation}
The condition of small change of the transverse momentum, which is used 
for derivation of all formulae  actually means the limitation on the length of the 
way passed: 
\begin{equation}
1 \le \frac{z}{{2\lambda p_0^2 }} \ll A,
\label{A}
\end{equation}
where
$A = \frac{{e\left\langle {p_{ \bot 0}^2 } \right\rangle ^2 }}{{2\pi ^2 \xi ^2 }};\frac{{\left\langle {p_{ \bot 0}^2 } \right\rangle ^3 }}{{2\pi ^2 \xi ^2 \Delta _\parallel  }};\frac{{p_{ \bot 0} }}{{\varepsilon \xi \pi }}$
for formulae (\ref{A1}), (\ref{A2}) and (\ref{A3}) respectively. The left unequality 
lets us to describe an adiabatic switching of a wave field. As in our case the field is 
switched instantly, the approximation of the adiabatic slow wave field change is fair 
at times greater than relaxation time. That is expression (6) for fair if the beam 
has passed a way greater than $2 p_0^2\lambda$. We shall emphasize that 
the condition (\ref{A}) comes from a method of calculation. The equation (\ref{eqmot}) is fair in 
the approximation of the given external field, when Langmuir (plasma) frequency of a beam is 
much less than frequency of the laser \cite{15}. It specifies relation between 
a beam transverse momentum variation and an energy variation. 
For a beam of charged particles with $P_0=100$ and divergence $\vartheta=10^{-3}$ and 
the laser with $\xi=10^{-2}$ the length of a way on which transverse emittance changes 
by 100 percents is about 80 cm, while the length of the relaxation is about 20cm. This 
example shows that very short cooling time is achievable by using 
resonant laser cooling.
Thus, the autoresonant cooling of beams of relativistic charges has some advantages. 
First, it is much faster and has shorter cooling time, than other cooling 
methods mentioned in the introduction. 
Second, this method  is applicable as to circular accelerators and storage rings 
as well as in linear accelerators.
Thirdly, though this method is most effective for beams with the 
Lorenz-factor about 100, is also can be used for cooling of GeV particles. 
The method allows to reach  cooling rates comparable to that 
described in \cite{10,11} by the use  of electromagnetic radiation sources 
with much lower  intensities. We notice that formulae which describe cooling depend on 
parameter $\xi$, which is easier be made of the value of about unity in long-wave 
radiation sources such as masers.

Fourth, there is a possibility of application of a longitudinal or 
transversal non-uniform magnetic field that enables scanning in case of a 
beam with large divergence.
\section{Dynamics of a circular accelerator beam in the presence of resonant laser cooling}
\subsection{Introduction} 
We consider in this section the influence of resonant laser cooling device on the 
beam dynamics of the circular accelerator. The beam dynamics in the presence 
of focusing forces of the accelerator is described by betatron oscillations around 
equilbrium orbit \cite{18}. During these oscillations the divergence (transverse momentum) 
and position of a particle varying periodically and the points of the maximum 
of one of them are the points of minimum of the other. So, the changing of the 
divergence of a particle by the resonant wave-beam interaction results in the 
appropriate changing of the particle position in respect to equilibrum orbit. This means, 
that the beam divergence reduction result in the accelerator beam size reduction. 
Therefore, the resonant beam-wave interaction result in decreasing of the 
betatron oscillation amplitude, i.e. in the accelerator beam cooling. This cooling 
is possible to apply in two regimes: fast cooling and slow cooling regimes. The first one 
is more suitable in case of electron beam due to a possibility to achieve very high 
cooling rates of light particles, such as electrons. This regime is necessary also 
for decreasing of the influence of the longitudinal magnetic field on the beam dynamics. 
This regime is considered in the first subsection. In the second subsection is 
considered the case of slow cooling more suitable for heavy particles such as 
protons and ions, but applicable to electrons too, under certain conditions . 
\subsection{Fast cooling regime} 
In this subsection we consider the case of significant changing, about unity or 
more times, of the beam angular divergence by resonant laser cooling on each 
turn on the accelerator orbit in the cooling device.This means that divergence 
decreased instantly and significantly at the cooling device. Such cooling regime is 
possible in case of cooling of electrons with $\gamma$ about 100 by $CO_2$ 
laser with energy about 0.5 kJ and magnetic field about 5-6 T. For describing of 
such process we consider the instantenous decreasing of the beam divergence 
$k$ times by passing through cooling device.
It is well known, that the transverse dynamics of the beam with emitance E in 
the focusing lattice of the accelerators is described by the ellipse \cite{18}
\begin{equation}
y^2+(\alpha y+\beta y')^2=\beta E
\label{ellipse}
\end {equation}
where y and y' denotes the transverse coordinate and the angular divergence, 
respestively. After passing the cooling device the transverse divergence in 
two transverse directions is decreased in accordance to (\ref{A1}). 
For effective cooling it is neccesary to place the cooling device in those 
points of the accelerator orbit, where  y' is maximal. Usually, these points are 
the centers of free spaces and in these points the relation \cite{18}
\begin{equation}
y'_{max} =\sqrt{E(1+\alpha^2)/\beta }
\end {equation}
take place. So, the decreasing of the angular divergence $k$ times means the 
decreasing of the transverse emmitance $k^2$ times. By using the relation \cite{18}
\begin{equation}
y_{max} =\sqrt{E\beta}
\end {equation}
one can find, that the mean square transverse size of the accelerator beam is 
decreased by factor $k$. The necessary cooling rate depend on laser intensity and 
pulse duration defined by period of one turn. Laser with intensity 5-6 MW decreases 
the transverse size of the electron beam with $\gamma=100$ approximately 2 times 
on each turn. In case $p_{\bot}<<1$ the longitudinal momentum change is negligible. 
Notice, that such decreasing of beam transverse size means decreasing of the 
beam emittance and increasing of the accelerator luminosity and of the beam lifetime 
in depend on factors which limited the beam lifetime.
\subsection{Slow cooling regime}
In this subsection we consider the case of small change of beam divergence 
by one pass through cooling device. This regime is more suitable in case 
of $\gamma$ about 1000 and in case of heavy particles: protons and ions. It is 
clear, that in this case the cooling time 
is very long and we can consider the square rout of the equation (\ref{basequat}) 
as a component of a force in the transverse directions \cite{18}
\begin{equation}
F_y  =  - a \xi \frac{{\pi \omega }}{{2p_0^2 \left\langle {p_ \bot } \right\rangle^3 }}\Delta _\parallel  (1 + \Delta _\parallel  )exp{( - {{\Delta _\parallel  } \mathord{\left/
 {\vphantom {{\Delta _\parallel  } {\left\langle {p_ \bot ^2 } \right\rangle }}} \right.
 \kern-\nulldelimiterspace} {2\left\langle {p_ \bot ^2 } \right\rangle }})} 
 \label{forcetr}
\end{equation}
where $a=L_m/l$, $L_m$ is the 
length of the circumference of the closed orbit, and $l$ is the length of the interaction 
way in the cooling device.
Notice, that in this case the cooling time is longer than in case of fast cooling 
considered in the previous subsection about 100 times, but much shorter than 
the heating time, which is about or longer than msec for electrons with $\gamma$ 
about 10000. So, even in this case and with laser of lower intensity the resonant laser  
cooling is very effective. The longitudinal cooling force can be found from the 
equation (\ref{longmom}) by assuming, that the average longitudinal momentum 
$p_0$ equals to equilibrum momentum $p_s$, which is defined, as well known, by 
the resonanticity of accelerating process.
\begin{equation}
F_\parallel =  - a\xi \frac{{\pi \omega }}{{2p_0 \left\langle {p_ \bot ^2 } \right\rangle ^2}}\Delta _\parallel  (1 + \Delta _\parallel  )exp{( - {{\Delta _\parallel  } \mathord{\left/
 {\vphantom {{\Delta _\parallel  } {\left\langle {p_ \bot ^2 } \right\rangle }}} \right.
 \kern-\nulldelimiterspace} {\left\langle {p_ \bot ^2 } \right\rangle }})} 
 \label{Flong}
\end{equation}
It is necesary to note, that the $p_0$ in the right hand size is a given function of time 
determined by acceleration. This force give an additional term in the longitudinal 
cooling rate determined by synchrotron radiation. Cooling time equals to \cite{18}
\begin{equation}
t_{cool} = \zeta^{-1}=(Fc/E_0)^{-1}
\end{equation}
where the $E_0$ is the equilibrum energy of the particle. Calculations show, that 
by using of solenoid with magnetic field about 5-10 T and laser with power about 5-10 kW, 
the cooling time is about few seconds for heavy particles with $\gamma=100$. 
The protons of this energy it is possible 
to cool by laser with wavelenght 10 mcm ($CO_2$ laser) but ions with maser with 
wavelength about or shorter 1mm. For comparison, the  electron cooling time 
of ions in the magnetic field about 5-10 T is about 10000 sec. 
So, the cooling time of resonant laser cooling of heavy particles is 10000 times 
shorter, than cooling time of electron cooling.
\section{Conclusion}
The resonant laser cooling, i.e. the laser cooling 
of particle beams in a longitudinal magnetic field has a number of advantages 
as compared with the newest laser \cite{10, 11} and undulator cooling 
methods. This method  provides much faster cooling given that all other 
conditions are the same. The cooling time of about few mcsec is achieveble in case 
of cooling of electrons in comparison to few msecs in other methods, for example, 
damper-magnets. In case of relativistic heavy particles the cooling time of about few 
seconds are achieveable in comparison to $10^4$ seconds cooling time of 
electron cooling. More realistic intensities of lasers (masers) are necessary in 
comparison to other laser cooling methods. Resonant laser 
cooling, if applied, can increase significantly the luminosity and lifetime of 
synchrotron accelerator beams, due to beam emittance decrease.  All this 
advantages come from the using of the stimulated radiation of the beam 
in the autoresonant laser for beam cooling. 
ACKNOWLEDGEMENTS. I am thankful to E. Laziev and M. Petrosian for interest 
to this work and useful discussions.

\end{twocolumn}
\end{flushleft}
\end{document}